\documentclass[prl,twocolumn]{revtex4}
\usepackage{epsfig}
\usepackage{amsmath}
\usepackage{graphicx}

\input{tcilatex}

\begin{document}

\title{Diameter-dependent electronic transport properties of
Au-catalyst/Ge-nanowire Schottky diodes}
\author{Fran\c{c}ois L\'{e}onard$^{1\ast }$, A. Alec Talin$^{1}$, B. S.
Swartzentruber$^{2}$, and S. T. Picraux$^{3}$}
\affiliation{$^{1}$Sandia National Laboratories, Livermore, California 94551}
\affiliation{$^{2}$Sandia National Laboratories, Albuquerque, New Mexico 87185}
\affiliation{$^{3}$Los Alamos National Laboratory, Los Alamos, New Mexico, 87545}
\date{\today }

\begin{abstract}
We present electronic transport measurements in individual
Au-catalyst/Ge-nanowire interfaces demonstrating the presence of a Schottky
barrier. Surprisingly, the small-bias conductance density increases with
decreasing diameter. Theoretical calculations suggest that this effect
arises because electron-hole recombination in the depletion region is the
dominant charge transport mechanism, with a diameter dependence of both the
depletion width and the electron-hole recombination time. The recombination
time is dominated by surface contributions and depends linearly on the
nanowire diameter.
\end{abstract}

\pacs{72.20.Jv, 73.63.Rt, 85.35.-p}
\maketitle

Semiconductor nanowires show promise as active elements in electronic \cite%
{dai,yang}, optoelectronic \cite{duan,samuelson}, and sensing \cite{lieber}
devices. Often, metal-catalyzed chemical vapor deposition is used to grow
the nanowires, with Au the most frequent catalyst metal. Typically, a small
hemispherical Au particle remains attached on the tip of the nanowires after
the growth is ceased \cite{kodambaka}. To date, the electrical nature of the
Au-catalyst/nanowire junction remains largely unknown. Yet, there are
several scientific and technological reasons to explore the electronic
transport characteristics of such contacts. For example, catalyst/nanowire
junctions offer a unique opportunity to examine how nanoscale dimensions
affect contact properties; and rectifying contacts to free-standing,
vertically oriented nanowires could prove useful in a number of applications
such as Schottky detectors and mixers.

Here, we use a microprobe inside of a scanning electron microscope (SEM) to
examine the charge injection at Au-catalyst/Ge-nanowire interfaces and
subsequent transport in the Ge nanowire. Our measurements indicate that this
interface is rectifying with a large Schottky barrier. Remarkably, the
current density \textit{increases} with decreasing nanowire diameter, in
contrast to common expectations. By modeling the nanowire electrostatics, we
show that this arises because the current is dominated by electron-hole
recombination in the depletion region, a contribution which is usually
negligible in bulk junctions, but is strongly enhanced in nanowires due to
the increased importance of surface recombination. Combining the modeling
results with the experimental data, we find that the recombination time
decreases as the nanowire diameter is decreased; a simple theory including
bulk and surface recombination explains this result.

The growth of Ge nanowires was performed in a cold wall chemical vapor
deposition system by the vapor-liquid-solid technique\cite{dailey} at a
temperature $\sim $375 $%
%TCIMACRO{\U{b0}}%
%BeginExpansion
{{}^\circ}%
%EndExpansion
$C and total pressure of 1.5 Torr. A 30\% GeH$_{4}$ precursor in H$_{2}$
along with 100 ppm PH$_{3}$ in H$_{2}$ as the source of the n-type dopant
was used with the gas flows set for a 1.2 x10$^{-3}$ P to Ge atom ratio. Au
colloids were used as the catalytic growth seeds on heavily doped (0.06%
%TCIMACRO{\U{3a9}}%
%BeginExpansion
$\Omega$%
%EndExpansion
-cm) n-type Ge (111) substrates with acidified deposition of the colloids
immediately prior to introduction into the growth chamber to achieve
predominately vertical nanowire growth. Based on previous results\cite%
{wang,tutuc}, we estimate that the carrier concentration is on the order of
10$^{18}-10^{19}$cm$^{-3}$.

\qquad The electrical measurements were carried out by contacting individual
vertically-oriented nanowires directly on the growth substrate using a
piezomotor controlled, Au-coated W probe retrofitted inside of a JEOL SEM
and connected to an Agilent semiconductor parameter analyzer\cite{statistics}%
. A SEM image of the probe near as-grown Ge nanowires is shown in the inset
of Fig. 1a. Most nanowires are $\sim $100 nm in height and have diameters
from 20-150 nm ($\pm $2 nm). Virtually all of the nanowires are capped with
the hemispherical Au catalyst nanoparticle.

\begin{figure}[h]
\includegraphics[width=7cm]{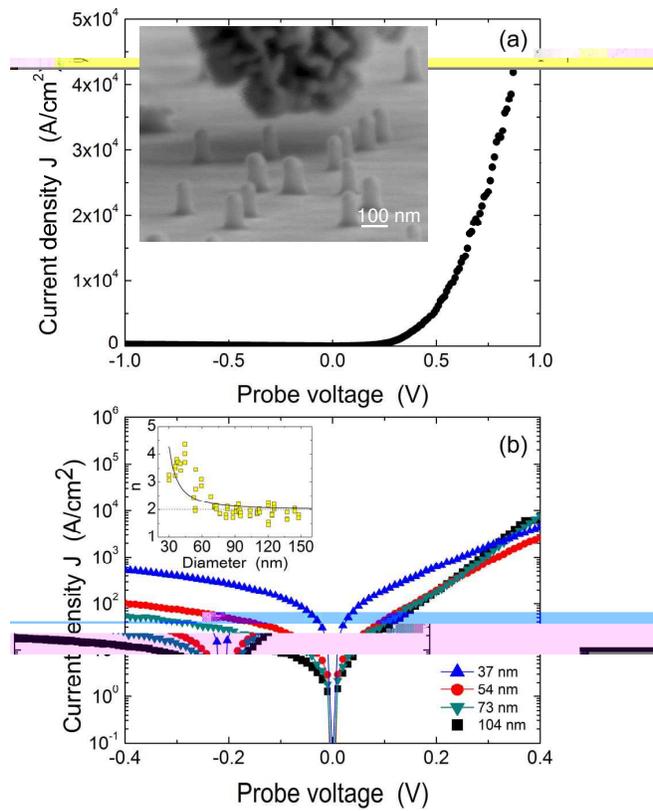}
\caption{(color online) Panel(a) shows current-voltage characteristics for a
Ge nanowire of 54 nm diameter. The inset is a SEM image of the Au-coated W
tip and several Ge nanowires. Panel (b) shows current-voltage curves on a
log scale, for four nanowires of different diameters. The inset shows the
ideality factor measured at forward bias as a function of nanowire diameter. 
}
\end{figure}

The current-voltage characteristics of a nanowire of 54 nm diameter is shown
in the main panel of Fig. 1a. The behavior is clearly rectifying, as
observed in almost all of the nanowires. This rectifying behavior is
consistent with that observed at bulk Au/Ge interfaces \cite{nishimura},
where a large Schottky barrier of 0.59 eV is present, and is essentially
independent of the type of metal because Ge has strong Fermi level pinning
close to the valence band. Our observations are also consistent with
atom-probe tomography measurements \cite{lauhon} and with high-resolution
transmission electron microscopy\cite{picraux} which indicated an abrupt
interface between the Au catalyst nanoparticle and the Ge nanowires. It is
tempting to analyze the rectifying behavior using conventional thermionic
emission theory to extract the Schottky barrier. However, a surprising
observation does not fit into existing theories of electronic transport
through Schottky barriers.

The observation, as shown in Fig. 1b, is that the small-bias conductance
density and reverse-bias current density are found to depend strongly on the
nanowire diameter, and in fact, \textit{increase} with decreasing diameter.
This is contrary to most models of transport in nanowires, where the
increased importance of surface scattering \textit{reduces} the small-bias
conductance density when the diameter is decreased. In addition, the
phenomenon cannot be explained based on a reduction of the effective
Schottky barrier height due to increased tunneling at smaller dimensions
because the depletion width actually increases with decreasing diameter, as
we will show below. To understand the experimental observations, and why the
current-voltage characteristics of Fig. 1b cannot be analyzed using
thermionic emission theory, we first consider the total current density in a
Schottky diode; this can be written as $J=J_{th}+J_{rn}+J_{rd}$, where $%
J_{th}$ is the thermionic current density, $J_{rn}$ is the electron-hole
recombination current density in the neutral region, and $J_{rd}$ is the
electron-hole recombination current density in the depletion region. The
first contribution is of the form \cite{sze} $J_{th}=A^{\ast }T^{2}\exp
\left( -\phi _{b}/kT\right) \left[ \exp \left( eV/kT\right) -1\right] $ with 
$A^{\ast }$ the Richardson constant, $\phi _{b}$ the Schottky barrier
height, $V$ the applied voltage, $k$ Boltzmann's constant, and $T$
temperature. In most diodes, this term dominates the behavior. In the
present case however, this contribution is negligible because of the large
Schottky barrier of 0.59 eV\cite{barrier}; indeed, the small-bias
conductance due to this term, $dJ/dV=\left( eA^{\ast }T^{2}/kT\right) \exp
\left( -\phi _{b}/kT\right) \sim 10^{-2}A/cm^{2}V$, is much smaller than
that measured experimentally $\left( \sim 1-20\text{ }A/cm^{2}V\right) $.
Similarly, the second term \cite{sze}, $J_{rn}=\left( n_{i}^{2}/N_{d}\right)
\left( ekT\mu /\tau \right) ^{1/2}\left[ \exp \left( eV/kT\right) -1\right] $
with $\mu $ the carrier mobility, $n_{i}$ the intrinsic carrier
concentration, $N_{d}$ the dopant concentration, and $\tau $ the
electron-hole recombination time, gives a negligible small-bias conductance
even with very low values of the recombination time. Thus, we are left with
the contribution \cite{sze} 
\begin{equation}
J_{rd}=J_{0}\left[ \exp \left( \frac{eV}{2kT}\right) -1\right]  \label{Jrd}
\end{equation}%
where $J_{0}$ depends on the depletion width and $\tau $. By analyzing the $%
J-V$ curves of the nanowires with the functional form $J\propto \left[ \exp
\left( eV/nkT\right) -1\right] $, we extract an ideality factor $n=2$ for
the larger nanowires as shown in the inset of Fig. 1b, in agreement with Eq. 
$\left( \ref{Jrd}\right) $. This suggests that electron-hole recombination
in the depletion region is the dominant transport mechanism in these
nanoscale contacts. (The reason for the larger values of $n$ at smaller
diameters will be discussed below.)

To further explore the role of electron-hole recombination, we performed
calculations of the small bias conductance of Ge nanowires. The small-bias
conductance density is written as\cite{sze}%
\begin{equation}
\left. \frac{dJ_{rd}}{dV}\right| _{V=0}=\frac{e}{2kT}\frac{1}{\tau \left(
d\right) }\int_{0}^{L}\frac{n_{i}^{2}}{n(z)+p(z)+2n_{i}}dz  \label{djdv}
\end{equation}%
where $n(z)$ and $p(z)$ are the electron and hole concentration at position $%
z$ along the nanowire of length $L$. In anticipation of the results below,
we have written the recombination time as being diameter dependent. The term
inside of the integral corresponds to the local recombination rate, assumed
to be independent of the radial coordinate, an assumption that is supported
by our three-dimensional calculations that we now describe.

To obtain the carrier concentrations $n(z)$ and $p(z)$, we calculate
self-consistently the charge and the electrostatic potential at zero bias
for the geometry of Fig. 2. There, a Ge nanowire of diameter $d$, length 100
nm, and dielectric constant $\varepsilon $ is making contact to a bulk metal
on one side and to the Ge substrate on the other side. A cylindrical
protrusion from the bulk metal with diameter and length equal to the
nanowire diameter was included to simulate the metallic nanoparticle. The
nanowire is surrounded by vacuum and is uniformly n-doped with dopant
concentration of 10$^{18}cm^{-3}$ (this choice of the dopant concentration
will be justified based on the results below). We numerically solve
Poisson's equation $\nabla \cdot \left[ \varepsilon \left( \mathbf{r}\right)
\nabla V\right] =-\rho \left( \mathbf{r}\right) $ using finite-difference,
with the spatially varying charge on the nanowire $\rho \left( \mathbf{r}%
\right) $ and with appropriate boundary conditions: at the metal/nanowire
interface, the strong Fermi level pinning fixes the electrostatic potential
to give a Schottky barrier of 0.59 eV, while at the nanowire/substrate
interface the potential is fixed to give charge neutrality for a doping of 10%
$^{18}cm^{-3}$. Boundary conditions are also applied at the nanowire surface
to produce the electric field discontinuity due to the different dielectric
constants of the nanowire and vacuum, and far away from the nanowire (1
micron in practice) the radial electric field vanishes. The local charge $%
\rho \left( \mathbf{r}\right) $ in the nanowire is calculated by integrating
the bulk density of states for Ge (shifted by the local electrostatic
potential) times the Fermi function at room temperature, $\rho \left( 
\mathbf{r}\right) =eN_{v}\int D\left[ E+eV\left( \mathbf{r}\right) \right]
f\left( E-E_{F}\right) dE$. We also consider the fact that Ge nanowire
surfaces contain a large density of surface states\cite{dai}. We thus
include an additional contribution to the charge given by $\rho _{s}\left( 
\mathbf{r}\right) =eN_{v}D_{0}e^{(r-d/2)/l}\int \left[ E_{F}+eV\left( 
\mathbf{r}\right) \right] f\left( E-E_{F}\right) dE$; this represents
surface states of uniform density in the Ge bandgap, with a neutrality level
at midgap; they decay exponentially into the nanowire with a decay length $%
l=0.3$ nm, and have a density at the surface $D_{0}=0.01$ states/eV/atom
(our conclusions are unchanged even if we vary this surface state density by
an order of magnitude). Finally, the electrostatic potential $V\left( 
\mathbf{r}\right) $ and the charge $\rho \left( \mathbf{r}\right) +\rho
_{s}\left( \mathbf{r}\right) $ are obtained self-consistently using Pulay
mixing on Poisson's equation and the expressions for $\rho $ and $\rho _{s}$.

\begin{figure}[h]
\includegraphics[width=8cm]{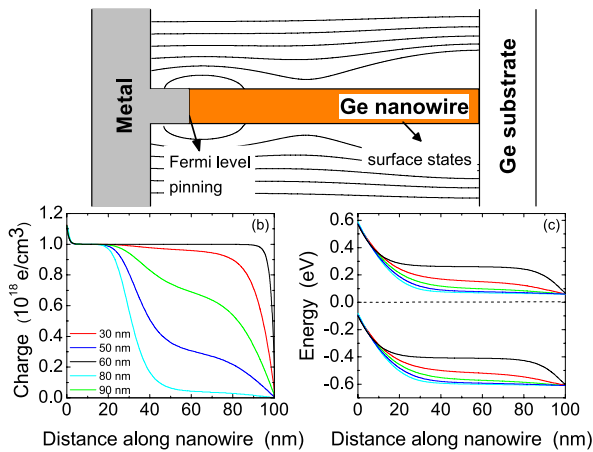}
\caption{(Color online) (a) Sketch of the system used for the numerical
calculations, see text for details. The calculated electric field lines are
shown for a nanowire of 30 nm diameter. (b) Calculated charge in the center
of the nanowire as a function of distance along the nanowire. (c) Calculated
band-bending. In (b) and (c) the curves from top to bottom correspond to
nanowire diameters of 30, 50, 60, 80, and 90 nm.}
\end{figure}
Figure 2 shows the calculated self-consistent charge and band-bending for
nanowires of different diameters. Near the metal/nanowire interface, the
nanowire is of p$^{+}$ character due to the position of the Fermi level
close to the valence band. A depletion region then extends over a length
that increases strongly as the nanowire diameter is decreased. As a
consequence, the total recombination current density in the depletion region
increases with decreasing diameter. To see if this behavior is sufficient to
explain our experimental data, we calculated the small bias conductance from
Eq. $\left( \ref{djdv}\right) $ assuming a \textit{fixed }recombination time
chosen to reproduce the large diameter values. When compared with the
experimental data in Fig. 3, the calculated small bias conductance with the
diameter-independent recombination time (dashed line) provides some increase
with decreasing diameter, but is insufficient to give the amount of measured
current and to reproduce the strong diameter dependence observed
experimentally. Thus, we conclude that the total recombination time must
depend on the nanowire diameter. To extract it from our experimental data,
we fit the calculated recombination current density to the experimental data
in Fig. 3 using a diameter-dependent recombination time $\tau ^{-1}\left(
d\right) =\tau _{bulk}^{-1}+a/d$; this gives excellent agreement with the
measurements as shown by the solid line in Fig. 3, with the value $%
a=7.85\times 10^{5}cm/s$.

To understand the dependence of the recombination time on nanowire diameter,
we consider an infinitely long nanowire into which carriers of density $%
n_{0} $ are injected initially. These carriers relax by diffusing through
the nanowire and recombining at the surface and in the bulk. Their time and
spatial dependence satisfy the diffusion equation \cite{mckelvey} $\partial
_{t}n=D\nabla ^{2}n-n/\tau _{bulk}$ where $D$ is the diffusion constant. At
the nanowire surface, electron-hole recombination with surface recombination
velocity $s$ takes place, giving the boundary condition \cite{mckelvey} $%
-D\left( \nabla n\cdot \widehat{r}\right) _{r=d/2}=sn(d/2,t)$. The solution
of these equations gives an exponential time decay of the carrier density,
with the smallest time constant $\tau ^{-1}=\tau _{bulk}^{-1}+\lambda \left(
s,D,d\right) $ where $\lambda $ satisfies%
\begin{equation}
D\lambda J_{1}\left( \lambda \frac{d}{2}\right) =sJ_{0}\left( \lambda \frac{d%
}{2}\right)  \label{J}
\end{equation}%
with $J_{\nu }$ a Bessel function of order $\nu $. For $s\ll D/d$ (which
applies to our data), one can use the small argument behavior of the Bessel
functions in Eq. $\left( \ref{J}\right) $ to get $\lambda =4s/d$ and 
\begin{equation}
\frac{1}{\tau }=\frac{1}{\tau _{bulk}}+\frac{4s}{d}.
\end{equation}%
Thus, the total recombination time is reduced by the presence of the surface
term. From our fit to the experimental data in Fig. 3, we extract $s\approx
2\times 10^{5}cm/s$; this value is consistent with that recently measured on
similar Ge nanowires using ultrafast time-resolved optical measurements \cite%
{rahid}. (The theory predicts $\tau ^{-1}\left( d\right) =\tau
_{bulk}^{-1}+3\pi D/d^{2}$ in the limit $s\gg D/d$, and this could be fitted
to the data of Fig. 3. However, this leads to a value for $D$ that is two
orders of magnitude lower than typical values for Ge.)

The dominance of electron-hole recombination in this system also explains
the large ideality factors measured at forward bias for the smaller
nanowires. Because the recombination current is essentially an integration
over the depletion region\cite{sze}, $J_{rd}\sim $ $W\left( V\right) \left[
\exp \left( eV/2kT\right) -1\right] $, where the bias dependence of $W$
arises because the built-in potential is reduced to $V_{bi}-V$ at forward
bias. In bulk diodes, this leads to a mild dependence $W_{bulk}(V)=\sqrt{%
2\varepsilon \left( V_{bi}-V\right) /eN_{d}}$, with little impact on the
ideality factor. However, for a nanowire $W$ depends exponentially on the
applied voltage, as we now discuss.

To calculate the depletion width, we approximate the charge distributions in
Fig. 2b as $\rho (r,z)=-eN_{d}\left[ 1-\theta \left( r-d/2\right) \right] $
for $0<z<W_{NW}$, where $W_{NW}$ is the nanowire depletion width. The
electrostatic potential \textit{at the center} of the wire is then $%
V(z)=\int_{0}^{d/2}\int_{0}^{W_{NW}}G\left( r=0,z;r^{\prime },z^{\prime
}\right) r^{\prime }dr^{\prime }dz^{\prime }$, where $G\left( r,z;r^{\prime
},z^{\prime }\right) $ is the electrostatic Green's function with the
boundary conditions that the derivative of the potential at the nanowire
surface be discontinuous by the ratio of dielectric constants of the
environment and nanowire; $G$ also includes the contribution from image
charges in the metal, and can be obtained using standard techniques. With
the requirement that $V(z)$ at the edge of the depletion layer give the
built-in voltage $V_{bi}$ this allows us to obtain (in the limit $%
W_{NW}/d\gg 1$)%
\begin{equation}
W_{NW}\approx d\exp \left( 8\frac{\varepsilon _{0}}{\varepsilon }\frac{%
W_{bulk}^{2}}{d^{2}}\right) .  \label{W}
\end{equation}%
This relationship is similar to that obtained for carbon nanotubes \cite%
{leonard}.

\begin{figure}[h]
\includegraphics[width=7cm]{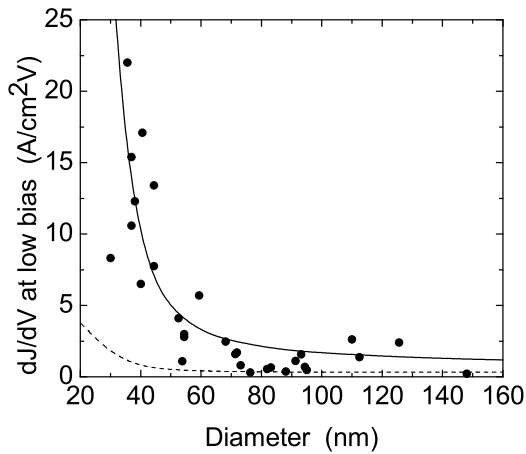}
\caption{Small-bias conductance density of the Au-nanoparticle/Ge-nanowire
interface as a function of the nanowire diameter. The dashed (solid) line is
calculated with a diameter-independent (-dependent) recombination time.}
\end{figure}
The exponential dependence of $W_{NW}$ explains the large forward bias
ideality factors shown in the inset of Fig. 1b. Indeed, we have $%
W_{NW}(V)\sim \exp \left( -16\varepsilon _{0}V/eN_{d}\right) $ giving $%
J_{rd}\sim \exp \left( \frac{eV}{n_{eff}kT}\right) $ with%
\begin{equation}
n_{eff}=2\left( 1-\frac{l^{2}}{d^{2}}\right) ^{-1}.  \label{neff}
\end{equation}%
Thus, the ideality factor increases as the nanowire diameter is reduced
below the length scale $l=\sqrt{32kT\varepsilon _{0}/e^{2}N_{d}}$. This
behavior can be tested by numerically fitting the data in the inset of Fig.
1b. As shown by the solid line, this provides a good description of the
measured ideality factor with the value $l\approx 22nm$; this compares
reasonably with the value of $7nm$ predicted by the expression for $l$. (We
note that at reverse bias, the depletion width should increase exponentially
with voltage, explaining the lack of saturation observed in our
measurements. But because $W_{NW}$ increases so rapidly, it can quickly
reach the finite length of the nanowire and start to deplete the substrate,
leading to a more complicated dependence on voltage. However, it is clear
from Fig. 1b that the reverse bias current density increases more strongly
as the diameter is reduced.) As mentioned earlier, the expected carrier
concentration is in the range 10$^{18}-10^{19}$ cm$^{-3}$; our results
suggest that the doping is actually close to the lower end of this range.
Indeed, for a doping of $10^{19}$ cm$^{-3}$, the calculated depletion width
is only 10 nm and does not vary over the experimental diameter ranges. Thus,
the condition $W_{NW}/d\gg 1$ is not satisfied, and the exponential
dependence of $W$ on voltage originating from Eq. $\left( \ref{W}\right) $
would not occur and the ideality factor would be independent of diameter.

To summarize, the unusual diameter-dependent electronic transport in this
system originates from several effects: Fermi level pinning at the Au/Ge
interface gives a large Schottky barrier and negligible thermionic current.
As a consequence, electron-hole recombination in the depletion region
dominates the current. This recombination current increases as the nanowire
diameter is reduced because the depletion width increases with decreasing
diameter, but mostly because the recombination time decreases due to the
added importance of surface recombination. At forward bias, the ideality
factor increases with decreasing diameter due to the electrostatics at
reduced dimensions. More generally, our results suggest that the electronic
transport properties of nanoscale contacts can differ significantly from
those of their bulk counterparts.

We thank D. Kienle for contributions to the self-consistent computational
approach. Work performed in part at the U.S. Department of Energy, Center
for Integrated Nanotechnologies, at Los Alamos National Laboratory and
Sandia National Laboratories.

$^{\ast }$email:fleonar@sandia.gov

\bigskip


\begin{references}
\bibitem{statistics} Two to four measurements were taken for each nanowire, by retracting the tip 
and re-contacting the nanowire. The I-V curves were essentially identical,
and the reported data is the average. This also implies that the SEM imaging
did not change the properties of the nanowires.
\bibitem{barrier} The Schottby barrier depends on 
atomic details of the Au/Ge interface; at the nanowire sizes considered
here, we expect a value similar to that of bulk interfaces.
\end{references}
\end{document}